\documentclass{article}

\usepackage{microtype}
\usepackage{graphicx}
\usepackage{subfigure}
\usepackage{booktabs} 

\usepackage{hyperref}

\usepackage[accepted]{icml2024}

\usepackage{amsmath}
\usepackage{amssymb}
\usepackage{mathtools}
\usepackage{amsthm}

\usepackage{cuted}
\usepackage{capt-of}
\usepackage{multirow}

\usepackage[capitalize,noabbrev]{cleveref}

\theoremstyle{plain}

\theoremstyle{definition}

\theoremstyle{remark}

\DeclareMathOperator*{\argmin}{arg\,min}
\usepackage{adjustbox}

\icmltitlerunning{Creative Text-to-Audio Generation via Synthesizer Programming}

\begin{document}

\twocolumn[
\icmltitle{Creative Text-to-Audio Generation via Synthesizer Programming}



\icmlsetsymbol{equal}{*}

\begin{icmlauthorlist}
\icmlauthor{Manuel Cherep}{equal,mit}
\icmlauthor{Nikhil Singh}{equal,mit}
\icmlauthor{Jessica Shand}{mit}
\end{icmlauthorlist}

\icmlaffiliation{mit}{Media Lab, Massachusetts Institute of Technology, Cambridge MA, USA}

\icmlcorrespondingauthor{Manuel Cherep}{mcherep@mit.edu}
\icmlcorrespondingauthor{Nikhil Singh}{nsingh1@mit.edu}

\icmlkeywords{Audio, Sound, Creative, Music, Text-to-Audio, Generation, Synthesis, Synthesizer}

\vskip 0.3in
]



\printAffiliationsAndNotice{\icmlEqualContribution} 

\begin{abstract}
Neural audio synthesis methods now allow specifying ideas in natural language. However, these methods produce results that cannot be easily tweaked, as they are based on large latent spaces and up to billions of uninterpretable parameters. We propose a text-to-audio generation method that leverages a virtual modular sound synthesizer with only 78 parameters. Synthesizers have long been used by skilled sound designers for media like music and film due to their flexibility and intuitive controls. Our method, \textit{CTAG}, iteratively updates a synthesizer's parameters to produce high-quality audio renderings of text prompts that can be easily inspected and tweaked. Sounds produced this way are also more abstract, capturing essential conceptual features over fine-grained acoustic details, akin to how simple sketches can vividly convey visual concepts. Our results show how \textit{CTAG} produces sounds that are distinctive, perceived as artistic, and yet similarly identifiable to recent neural audio synthesis models, positioning it as a valuable and complementary tool.\footnote{\href{https://ctag.media.mit.edu/}{ctag.media.mit.edu}}
\end{abstract}

\section{Introduction}
\begin{quote}
    \textit{``Of course, bubbles don't make sound, but this is the magic of sound design...you can create the concept of a sound and it seems real.''}
    \\
    \vspace{2mm}
    \hfill --- Suzanne Ciani
\end{quote}
In creative sound design, realism isn't everything. In the late 1970s, composer Suzanne Ciani famously demonstrated this principle with her iconic \textit{Coca Cola pop and pour} sound effect. This sound, which has become synonymous with the refreshing experience of opening a soda, was not recorded from an actual soda bottle, but skillfully crafted using a Buchla synthesizer. Ciani's work illustrates the immense power of abstraction in auditory representation, where the essence of a concept can be expressed without mimicking real-world acoustic details, while achieving greater impact.

This approach extends beyond single examples into the domain of procedural sound design: creating sounds algorithmically using parameters that can be manipulated to achieve desired sonic effects. By applying procedural techniques, sound designers can often transcend what's physically plausible to obtain by recording real-world events. These methods can lead to highly evocative and expressive sounds in music, film, video games, advertising, product design, and other media.

\begin{figure}[t]
\centering
\includegraphics[trim={3cm 1.5cm 3cm 1cm},clip, width=\columnwidth]{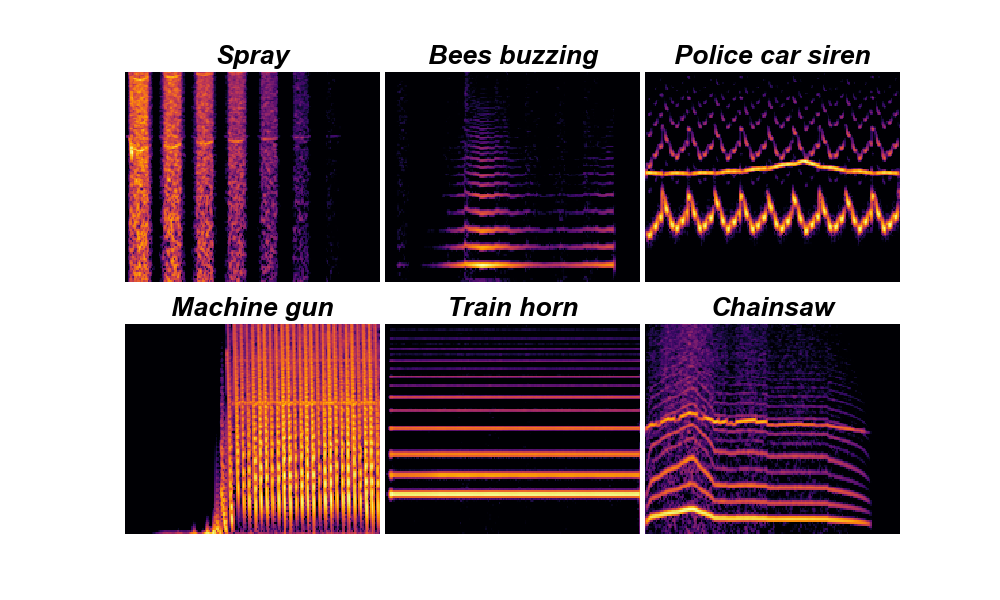}
\caption{\textit{CTAG} leverages a virtual modular synthesizer to generate sounds capturing the semantics of user-provided text prompts in a sketch-like way, rather than being acoustically literal. Spectrograms of auditory outputs corresponding to six text prompts showcase the range of sounds this approach can yield, accompanied by a fully interpretable and controllable parameter space.}
\label{fig:examples}
\end{figure}

Neural audio synthesis methods have transformed the state of sound design, enabling specifying sound ideas using intuitive inputs like textual prompts. However, there remains unrealized potential in integrating expressive sound design principles into neural audio synthesis. Current techniques prioritize acoustic recreation and end-to-end application, often overlooking creative possibilities for evoking emotions or concepts, and interactive aspects like manipulating, iterating, and interpolating between sounds. While recent advances showcase remarkable capabilities in replicating real-world sounds, this emphasis can limit the creative palette and expressive potential of generated audio. We propose a method to bridge this gap.

Overall, this work contributes:
\begin{itemize}
    \item A novel method that integrates a virtual modular synthesizer with a pretrained audio-language model for generating sounds that resonate with human intuition without being literal representations.
    \item A lightweight, fully interpretable, and controllable synthesizer resulting from our approach, allowing for easy inspection and tweaking for creative purposes.
    \item Extensive experiments evaluating different approaches to solving this problem, varying optimization algorithms, sound durations, and synthesis architectures.
    \item Qualitative and quantitative results that highlight how sounds from our method have distinct features from those produced by other neural audio generators, while still being identified at similar rates. We conduct a user study as a gold standard evaluation, given the novelty of the task, which shows the identifiability and potential artistic value of \textit{CTAG}'s sounds.
    \item Examples of this approach generating several datasets of sounds with their synthesis parameters, and interpolating between different sounds in the parameter space.
\end{itemize}

We will open-source our approach, both to provide a tool for novices and experts alike to realize their ideas, as well as to provoke future audio generation paradigms that recognize abstraction as an important factor for creative expression.

\section{Related Work}
\subsection{Sound Synthesis}
Neural audio synthesis consists of two main strands: approaches that generate audio waveforms directly in the time domain, and those that do so in the frequency domain. WaveNet \cite{oord2016wavenet} notably introduced an autoregressive approach to audio synthesis by predicting one sample at a time. This slow iterative sampling approach, later refined in SampleRNN \cite{mehri2016samplernn} and WaveRNN \cite{kalchbrenner2018efficient}, reflects the sequential nature of audio data, in contrast to images wherein GANs with global latent conditioning and efficient parallel sampling quickly became a dominant method for synthesis. Later, WaveGAN \cite{donahue2018adversarial} and GANSynth \cite{engel2019gansynth} demonstrated that GANs could in fact be used to synthesize locally-coherent audio, outperforming sequential models' speed by several orders of magnitude while maintaining a focus on high-fidelity, natural-sounding audio.

A third strand of so-called \textit{oscillator} models, largely propelled by Differentiable Digital Signal Processing (DDSP) \cite{engel2020ddsp} is physically and perceptually motivated by the rich history of synthesis and signal processing techniques. Our approach is motivated by this direction, but relies on a simple synthesizer architecture, CLAP~\cite{wu2023large}, for text-conditioning, and gradient-free optimization to provide a simple, training-free solution.

\subsection{Language-Sound Correspondence}
Advances in multi-modal sound-language models have been partly motivated by CLIP~\cite{radford2021learning} for images. Wav2CLIP \cite{wu2022wav2clip} builds directly onto CLIP by adding an audio encoder, and VQGAN+CLIP \cite{crowson2022vqgan} generates and edits images guided by text prompts. Audio representation models, such as Microsoft's CLAP \cite{elizalde2023clap} and LAION-CLAP \cite{wu2023large}, emulate CLIP's approach by using contrastive learning on audio-text pairs. We use LAION-CLAP as our audio-language model in this work.

Other recent approaches cast audio generation as a language modeling task. AudioGen~\cite{kreuk2022audiogen} is an autoregressive model conditioned on text inputs. AudioLM~\cite{borsos2023audiolm} uses a multi-stage Transformer-based language model. WavJourney~\cite{liu2023wavjourney} uses text instructions to create scripts, which are then used for compositional audio creation. Make-An-Audio 1 and 2~\cite{huang2023make,huang2023make2} offer text-to-audio synthesis with prompt-enhanced diffusion models, using CLAP to map text to latent representations with a spectrogram autoencoder. AudioLDM~\cite{liu2023audioldm} learns continuous audio representations from CLAP latents and can perform text-guided audio manipulations. We compare to two state-of-the-art solutions, namely \textit{AudioGen} and \textit{AudioLDM}, in our experiments. Our goals differ significantly from those of these models, as we seek to generate abstract yet high-quality sounds, rather than literal recording-like renditions.

\subsection{Abstract Synthesis}
Visual sketching offers an intuitive analog to abstract sound synthesis. Minimal representations like monochromatic line drawings might use only straight lines and curves with no additional shading or color. These renderings are non-photorealistic; they evocatively convey meaning while emphasizing a subject's essence over its real-world presentation. They can also reveal insights about a subject's underlying geometry, proportions, and symbolism that may be obscured in more realistic depictions.

The problem of computing recognizable and insightful abstract renderings has seen more progress in the visual than the audio domain. CLIPasso \cite{vinker2022clipasso} leverages CLIP to distill semantic meanings from images and sketches alike and thereby guide text-to-image generation, varying the number of strokes according to the desired level of abstraction. CLIPTexture \cite{song2022cliptexture} enables a user to manipulate a simple sketch or layout through textual descriptions. CLIPVG \cite{song2023clipvg} follows the same progressive optimization approach, but performs image manipulation using vector graphics rather than pixels. ES-CLIP \cite{tian2022modern} tackles the problem via evolution strategies, generating configurations of colored triangles on a canvas, then assessing their fitness for further iteration. We were inspired by this approach, though we rely on the well-established, easily interpretable, and tweakable paradigm of modular synthesis.

In the auditory domain, the Sound Sketchpad \cite{singh2021sound} combines sounds together using audio-visual sketches, and the SkAT-VG project \cite{rocchesso2015sketching} applies vocal and gestural manipulation as natural sketching tools. In our approach, we focus on language input, and synthesis rather than the composition of existing sounds.

\subsection{Interpretable and Controllable Synthesis}
Interpretability and controllability of results is essential to human-machine co-creation, in which it is often desirable to closely examine, understand, and fine-tune an artifact. For creative sound design using neural synthesis methods, it can be impossible to retrace decisions made by a complex neural synthesis model en route to synthesizing an output. The model may also not provide any opportunity to iteratively refine the output. Some prior work~\cite{NEURIPS2022_f13ceb1b} highlights the potential of program synthesis for interpretability in sequence data, including music. Some neural synthesis models integrate techniques like timbre-regularization~\cite{Esling2018BridgingAA} to bridge powerful synthesis methods with perceptually-motivated organization of latent spaces. By contrast, our approach offers a fully interpretable and controllable parameter space without requiring us to develop additional neural infrastructure.

\subsection{The Synthesizer Programming Problem}
Despite the near-ubiquitous presence of synthesized sound in modern music, synthesizer programming---that is, the act of creating new sounds through careful analysis and modulation of synthesizer parameters---is a complex task that can often impede the creative process, if not bar entry entirely. In particular, the conceptual disconnect between parameter settings and the associated auditory output \cite{shier2021synthesizer} makes synthesizer programming especially non-intuitive without special training. Recent work has investigated techniques for inverse synthesis---given a target sound, infer the parameter setting that will emulate the sound to the closest extent possible---on both musical sounds \cite{yee2018automatic} and real-world sounds, such as animal vocalizations \cite{hagiwara2022modeling}, including deep learning methods to learn invertible mappings~\cite{Esling2020FlowSynthSC}. However, this task still requires a specific audio clip to start. We provide text-to-parameter inference to bridge this gap, generalizing beyond specific audio files to broader semantic notions of arbitrary sounds.

\section{Methods}

\begin{figure*}[t]
\centering
\includegraphics[width=\textwidth]{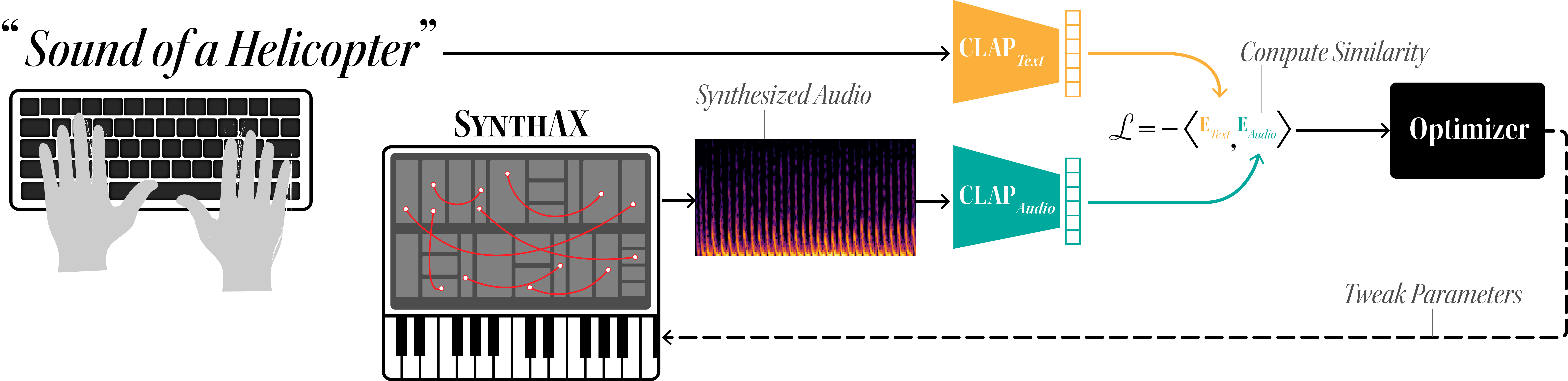}
\caption{High-level overview: we use the LAION-CLAP model~\citep{wu2023large} to compute the similarity between a user-provided text prompt and \textsc{SynthAX}'s~\citep{cherep2023synthax} output. The optimization procedure iteratively adjusts the parameter settings.}
\label{fig:system_diagram}
\end{figure*}

Our methodology hinges on three pillars: a synthesizer, implemented via \textsc{SynthAX} \cite{cherep2023synthax}, gradient-free optimization methods, implemented via the Evosax \cite{lange2023evosax} evolutionary optimization library, and an objective function based on the LAION-CLAP \cite{wu2023large} model, which we use to estimate semantic alignment between the synthesized audio and its corresponding text prompt (see \Cref{fig:system_diagram} for an overview of the pipeline).

\subsection{Synthesizer}
We use a simple synthesizer implementation available in \textsc{SynthAX}, a fast modular synthesizer written in JAX~\cite{bradbury2018jax}. We specifically use the \textit{Voice} synthesizer architecture, adapted from \textit{torchsynth}~\cite{turian2021one}, which has already been used for programmatic resynthesis of sounds~\cite{hagiwara2022modeling}. It consists of 78 parameters for a monophonic keyboard, two low-frequency oscillators (LFOs), six ADSR envelopes, a sine voltage-controlled oscillator (VCO), a square-saw VCO, a noise generator, voltage-controlled amplifiers (VCAs), a modulation mixer and an audio mixer. All parameters are initialized uniformly, $\theta_i~\sim~U(0, 1)$.

In addition to this architecture, we evaluate the following variants in increasing order of architectural complexity:
\begin{itemize}
    \item \textit{ShapedNoise}: An 18 parameter synthesizer consisting of a noise generator, and two control elements to shape the noise amplitude over time: an ADSR envelope, and a low-frequency oscillator (LFO). These are combined into a modulation signal through a modulation matrix, which itself has learnable weights for this combination.
    \item \textit{OneOsc}: A 23 parameter synthesizer consisting of a sine wave voltage-controlled oscillator (VCO), and the same two control elements as above. These elements are combined into two signals through a modulation matrix, one each for frequency and amplitude.
    \item \textit{NoLFO}: A 29 parameter two-VCO synthesizer, where one is a sine wave oscillator and the other is a square-saw wave oscillator with a ``shape'' parameter which controls the degree of ``square-ness'' vs. ``saw-ness''. This synthesizer has no LFO components, all modulation is conducted by two ADSR envelopes combined into four separate modulation signals (pitch and amplitude controls for each of the two VCOs).
    \item \textit{NoNoise}: A 51 parameter synthesizer with two VCOs (as before), and a more complex modulation structure. Here, there is a single LFO, but there are additional ADSRs to modulate the frequency and amplitude of this LFO. The modulated LFO and two ADSR envelopes comprise the inputs to the modulation matrix.
    \item \textit{Voice+FM}: A 130 parameter synthesizer which adds a frequency modulation (FM) component to the original \textit{Voice} architecture.
\end{itemize}

For reference, an ADSR envelope is a piecewise control signal consisting of linear or exponential segments: \textbf{A}ttack, \textbf{D}ecay, and \textbf{R}elease, which specify the duration of each envelope segment. The \textbf{S}ustain parameter is the level of the control signal after the decay phase. An LFO is an oscillator whose frequencies are typically lower than audible frequencies, i.e. below 20-40 Hz. These are used for periodic control of synthesis parameters.

In all our experiments, the synthesizer has a control rate of 480 Hz and the audio is generated in batches at a sample rate of 48 kHz. This sample rate is much higher than that commonly used for neural audio synthesis systems (often 16 kHz) and therefore admits much more high-frequency content to be generated.

\subsection{Optimization}
\label{ssec:optimization}

\begin{algorithm}[!htb]
\begin{algorithmic}
\REQUIRE Text prompt $p$
\REQUIRE Population/batch size $N$  
\REQUIRE Iterations $M$

\medskip 
\STATE \textbf{Components:}
\STATE CLAP text embedding model $C_t(p) \rightarrow E^p$
\STATE SynthAX synthesizer $S(\Theta) \rightarrow X^a$
\STATE CLAP audio embedding model $C_a(X^a) \rightarrow E^{X^a}$
\STATE Optimization Strategy: $O$

\medskip 
\STATE \textbf{Initialize:} 
\STATE Synthesis parameters $\Theta = \{\theta_1,\ldots,\theta_N\}, \theta_i \sim U(0, 1)$  
\STATE Flattened parameters $\Theta_f \in [0, 1]^{N \times d} = Flatten(\Theta)$  

\medskip 
\FOR{$i=1$ {\bfseries to} $M$}
\STATE $\Theta_{f_\textrm{new}} \gets O_{\textrm{ask}}(\Theta)$ \hfill \textit{Generate candidates}
\STATE $\Theta_{\textrm{new}} \gets \textrm{Reshape}(\Theta_{f_\textrm{new}})$ \hfill \textit{Reshape}
\STATE $X^a \gets S(\Theta_{\textrm{new}})$ \hfill \textit{Synthesize audio}
\STATE $E^{X^a} \gets C_a(X^a)$ \hfill \textit{Get audio embeddings}
\STATE $F \gets -E^{X^a}{E^p}^T$ \hfill \textit{Compute fitness}
\STATE $O_{\textrm{tell}}(\Theta_{\textrm{new}}, F)$ \hfill \textit{Update optimizer state}
\STATE $\Theta \gets \Theta_{\textrm{new}}$
\ENDFOR

\medskip  
\STATE $\theta^* = \argmin_\theta F$ \hfill \textit{Select optimal parameters} 
\end{algorithmic}
\caption{Our optimization procedure for producing sounds in \textit{CTAG}. Note: $d$ is the number of parameters of the synthesizer $S$; for simplicity we omit batches.}
\label{alg:ctag}
\end{algorithm}

During initial experiments, we found the gradients of our differentiable synthesizer to be highly unstable. This instability hindered optimization performance even after attempting mitigation strategies. Recent works in abstract visual synthesis have shown that non-gradient methods can achieve state-of-the-art results without relying on gradient information~\cite{tian2022modern}. Given these findings, we decided to explore non-gradient approaches which are more suitable for our synthesizer's instability and have demonstrated effectiveness for this task. Focusing efforts here allowed us to sidestep gradient issues while leveraging successful techniques from related synthesis domains.

We experimented with several non-gradient optimization algorithms, using implementations from Evosax~\cite{lange2023evosax}. Specifically, we examined simple baselines like random search and a simple genetic algorithm~\cite{such2017deep}, well-known methods like CMA-ES~\cite{hansen2001completely} and Particle Swarm Optimization~\cite{kennedy1995particle}, and state-of-the-art methods like Learned and Discovered Evolution Strategies~\cite{lange2023discovering}. For each algorithm, we first tuned hyperparameters using Bayesian optimization via the Adaptive Experimentation (AX) platform \cite{bakshy2018ae}. We tuned for 50 trials on the ESC-10 dataset, a subset of ESC-50~\cite{Piczak2015ESCDF}. Note that the hyperparameter tuning uses no privileged information and can easily be applied downstream on new prompt sets to maximize the performance.

The optimization procedure is specified in \Cref{alg:ctag}.

\subsection{Objective Function}
We use LAION-CLAP \cite{wu2023large} with an HTSAT-based audio encoder~\cite{chen2022hts} and a RoBERTa-based text encoder~\cite{liu2019roberta}. We used the \textit{audioset-best} checkpoint for general audio less than 10 seconds long.

The encoders process the audio data $X_i^a$ in batches of size $\mathcal{B}$ where $\mathcal{B}$ corresponds to the optimizer's population size, along with a prompt $p$. Note that $(X_i^a, p)$ is one particular pair of synthesized audio with input text prompt. We extract the audio embeddings $E_\mathcal{B}^a \in \mathbb{R}^{\mathcal{B} \times 512}$ and the text embeddings $E^p \in \mathbb{R}^{1 \times 512}$ with the encoders and use them to calculate the similarity score between a batch of audio data and a specific prompt. 

\begin{equation}
\label{eq:synthesizer}
    X_i^a = \mathcal{S}(\theta_i)
\end{equation}

\begin{equation}
\label{eq:minimization}
    \theta^* = \argmin_{\theta}\ - E_i^{\mathcal{S}(\theta_i)} {E^p}^T
\end{equation}

\Cref{eq:synthesizer} shows how the synthesizer $\mathcal{S}$ takes parameters $\theta_i$ and produces a sound (in practice, this is done batched). Then \Cref{eq:minimization} 
formulates the optimization problem to optimize the similarity score between each audio in the batch and one given text prompt using their corresponding embeddings.

\subsection{Evaluation Metrics}
\label{sec:eval_metrics}
Since we propose a novel synthesis task without existing evaluation metrics, we devise a principled evaluation suite that allows us to quantitatively assess our contributions, in addition to qualitatively reviewing synthesized examples.

\paragraph{Classification Experiments}
To determine whether our generated sounds are more abstract than neural synthesis methods, we compared results on pretrained classifiers with sounds generated from their class labels. Lower scores can indicate a distribution shift from real audio, despite explicitly optimizing for similarity to the label. We complement with human listener ratings.

Without a perfect synthesis engine, any methods to generate sound will introduce a distribution shift from real audio. In our case, there is a deliberate domain shift to abstract audio. We evaluate on two well-known datasets. The first is ESC-50, a 50-class canonical environmental sound classification dataset~\cite{Piczak2015ESCDF}. The second is a subset of AudioSet~\cite{Gemmeke2017AudioSA}; the full ontology of classes is very large (over 500). We consider classes from ``sounds of things'' given that this category contains the most sub-classes and sub-selected the top 50 classes by number of annotations, removing duplicates or equivalent classes. We use a pretrained Audio Spectrogram Transformer (AST) model for AudioSet-50, and fine-tune an AST for ESC-50 classification~\cite{Gong2021ASTAS}. When evaluating on AudioSet-50, we mask the remaining logits to effectively make it a 50-class classifier.

\paragraph{Synthesis Quality}
A significant benefit of our approach is synthesizing clean audio using signal generators while keeping attributes like sample rate flexible. We find synthesized sounds also often exaggerate aspects of the prompts, resulting in large variations in acoustic properties over time. Evaluating audio quality reference-free is challenging, so we examine acoustic features that correlate with these aspects (such as high-frequency content and spectral variation).

\begin{figure*}[!htb]
    \centering
    \includegraphics[width=\textwidth]{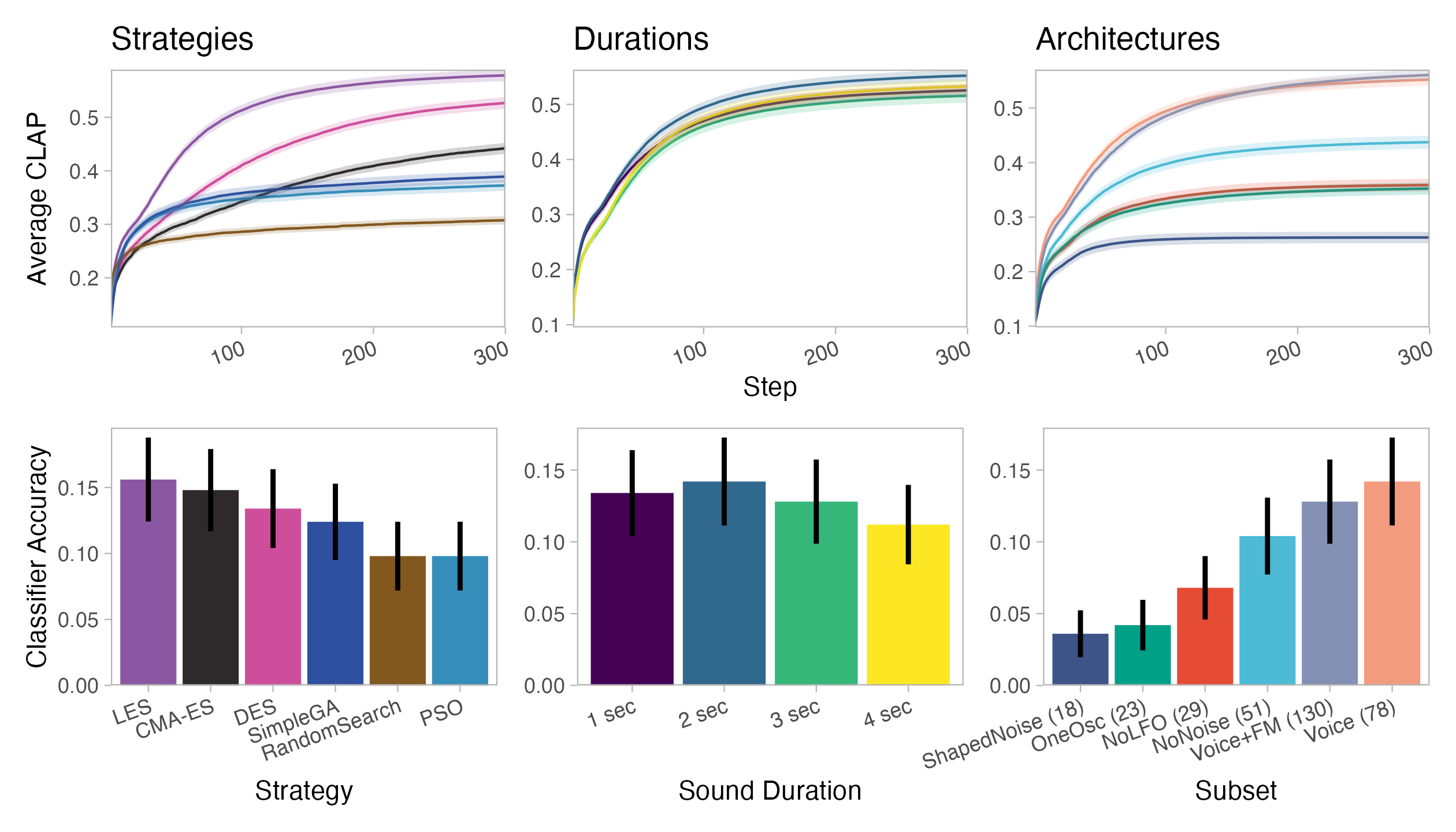}
    \caption{Results from our ablation study; all experiments are conducted with ESC-50. \textbf{(Top)} CLAP maximization curves, averaged across 10 iterations for each of the 50 prompts. Colored bands show 95\% confidence intervals. \textbf{(Bottom)} Classification accuracy, with error bars showing 95\% confidence intervals. Top and bottom plots share colors. \textbf{(Left)} Performance of different algorithms, with hyperparameters tuned on ESC-10. LES is strongest in both optimization and downstream classification. \textbf{(Center)} Different sound durations; we find 2 seconds to be strongest. \textbf{(Right)} Impact of synthesizer architecture, finding strongest results from the \textit{Voice} model. Parameter counts are given in parenthesis, such as (78) for \textit{Voice}.}
    \label{fig:results}
\end{figure*}

\paragraph{User Study}
We conduct a listening test with human evaluators. We ask them to classify sounds, rate their confidence, and rate sounds along a scale from realistic portrayal to artistic interpretation. This offers us the most direct signal of our abstraction-related goal. We share details on this study in the next subsection. We compared against the recent neural generation methods \textit{AudioLDM} \cite{liu2023audioldm} and \textit{AudioGen} \cite{kreuk2022audiogen}.

From our 50-prompt subset of AudioSet~\cite{Gemmeke2017AudioSA} classes, we randomly selected 10 for this study. We used text embeddings of the labels with a facility location submodular optimization algorithm from the apricot package~\cite{schreiber2020apricot} to select a modest-sized semantically representative subset. Within each prompt, we randomly sampled two of 10 available \textit{CTAG} sounds. The prompts were: \textit{Truck air brake}, \textit{Water tap}, \textit{Train horn}, \textit{Motorcycle}, \textit{Microwave oven}, \textit{Liquid slosh}, \textit{Chainsaw}, \textit{Airplane}, \textit{Bicycle bell}, and \textit{Machine gun}. For \textit{AudioLDM} and \textit{AudioGen}, we used their default parameters to generate two sounds per prompt.

This study was determined to be exempt by our institution's IRB. Each participant rated 60 sounds (20 per method) in random order. To examine category-level recognition, participants were asked to select a category given a list of options and rate their confidence. To determine whether our generated sounds were perceived as (abstract) artistic interpretations, we posed the question: ``Would you associate this sound more with a realistic portrayal or an artistic interpretation of the label that you selected?'' with options on a scale from 1 (realistic portrayal) to 5 (artistic interpretation). We modeled participant responses with mixed-effects logistic and linear regression models and post-hoc contrasts.

\section{Results}
\subsection{Ablation Studies}
\Cref{fig:results} shows results from our ablation studies, including, from left to right, (1) optimization algorithms with tuned hyperparameters, (2) sound durations, and (3) synthesis architectures. Overall, we observe that the LES algorithm significantly outperforms our other options within the computation budget of 300 iterations (more than needed for several prompts). This experiment was conducted with 2-second long sounds, which we observe in the \textit{Durations} experiment to yield a higher overall CLAP score and classification accuracy than 1, 3, or 4-second long generations. Finally, we see that the \textit{Voice} architecture yields the best results, offering a balance of flexibility in its parameters and modular structure, as well as ease of optimization. However, we note that expanding to larger architectures like VoiceFM could be useful for future work to explore, with more work on the optimization strategy to obtain the best results.

Based on these results, we conduct all additional experiments discussed with the LES optimizer, 2-second sounds, and the \textit{Voice} architecture. We conducted a full hyperparameter tuning run with 50 trials of all ESC-50 prompts to obtain the final optimization hyperparameters.

\subsection{Qualitative Results}

\subsubsection{Examples}

\Cref{fig:examples} shows spectrograms of sounds---given in the supplementary material---corresponding to six text prompts. The ``spray'' shows bands of noisy bursts, reflecting the short, sharp sound of aerosol being expelled. The ``bees buzzing'' presents a band of low to high frequencies, encapsulating the vibrant hum of a bee. The ``police car siren'' is characterized by high-frequency oscillations that sharply rise and fall. The ``machine gun'' reveals rapid, staccato bursts of energy across a broad frequency range. The ``train horn'' displays horizontal bands across mid to high frequencies, illustrating the horn's fundamental tone and its partials. Lastly, the ``chainsaw'' spectrogram is dominated by intense, continuous mid-range frequencies, punctuated by peaks corresponding to the engine's roaring and cutting action.

\subsubsection{Interpolation}
\begin{figure}[ht]
\centering
\includegraphics[width=\columnwidth]{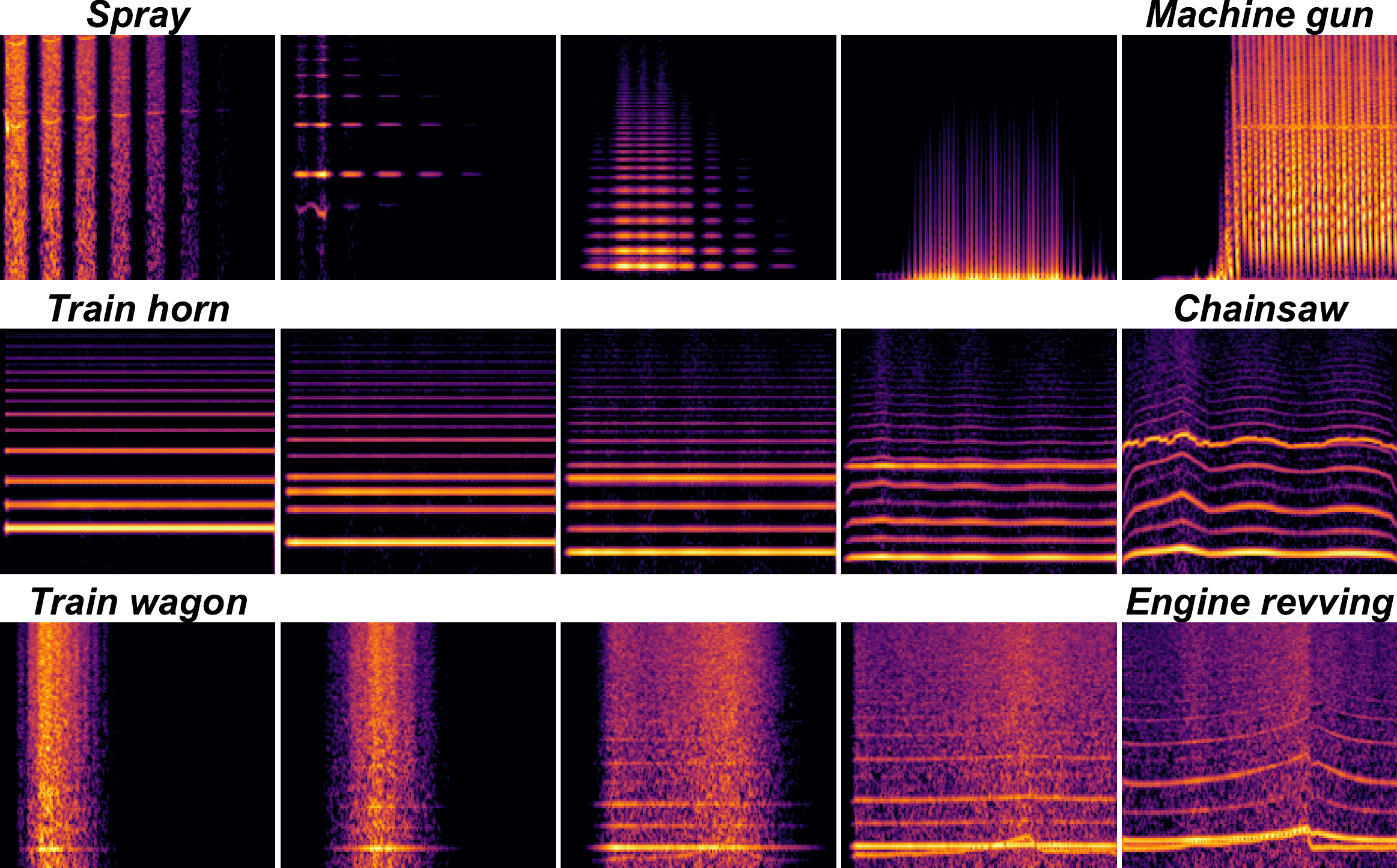}
\caption{Spectrogram series as the result of linear interpolation of the synthesizer's parameters (1) from ``Spray'' (left) to ``Machine gun'' (right), (2) from ``Train horn'' to ``Chainsaw'', and (3) from ``Train wagon'' to ``Engine revving''. Each spectrogram in the sequence represents a step in the interpolation, highlighting the systematic shift in acoustic properties.}
\label{fig:interpolation}
\end{figure}

In sound synthesis, interpretable parameters offer a unique opportunity for deeper insight. Our method provides a fixed set of parameters that possess this property---a salient distinction from contemporary models equipped with high-dimensional latent spaces. This interpretability extends to interpolation between parameters of distinct sounds, granting auditory access to intermediate acoustical transitions. In \Cref{fig:interpolation}, we present a systematic series of spectrograms between pairs of prompts: (1) ``Spray'' to ``Machine gun'', (2) ``Train horn'' to ``Chainsaw'', and (3) ``Train wagon'' to ``Engine revving,'' with three intermediary steps linearly interpolated. This discernible gradation corroborates the capacity of our parameter space to retain congruence.

\subsection{Classification Results}
\label{sec:classification_results}

Results are shown in \Cref{tab:results_c}. On AudioSet-50, our results are higher than \textit{AudioLDM}. On ESC-50, the classifier recognizes \textit{CTAG}'s sounds the least, showcasing the distribution shift from its training on realistic sounds. We experimented with constructing concise and descriptive prompts from each sound class from both ESC-50 and AudioSet-50. We used GPT-4 \cite{OpenAI2023GPT4TR} to automatically produce caption-style prompts. We also tried a simple template (i.e. ``Sound of a/an ...'') to compare. \Cref{tab:results_c} also shows results for these template (\textit{CTAG+T}) and caption-style prompts (\textit{CTAG+C}). Introducing such strategies does not appear to greatly influence classifier identification. However, in a few cases, we observed the elaborated prompts helped to produce qualitatively more accurate results. Overall, \textit{CTAG} sounds are classified correctly significantly higher than chance, and competitively with \textit{AudioLDM}.

\subsection{Synthesis Quality and Variation}
Evaluating the quality of generated examples is challenging for two reasons. First, we lack auditory references to compare against, as we generate from text directly and never use text-audio reference pairs. Most audio quality metrics are reference-based. Second, distance-based metrics such as FAD will likely be confounded by realism. \textit{CTAG}'s sounds are high-quality in that they can be generated at high sample rates and are free of noise or artifacts owing to real-world recording environments or neural synthesis.

To evaluate, we use auditory descriptors (implemented using Essentia~\cite{Bogdanov2013EssentiaAA}) that are plausible correlates of these notions of quality, shown in \Cref{tab:results_q}. Spectral complexity highlights the presence of more peaks, signaling diversity in the timbral components, while flux shows greater variation of timbre over time for \textit{CTAG} compared with other methods. Following these, HFC (high-frequency content), spectral rolloff, and spectral centroid provide signals of ``brightness'' or high-frequency presence in the sounds. All of these results show our method's ability to introduce high-frequency content into generated sounds, likely in part due to the higher sample rate we use.

\begin{table*}[!htb]
    \centering
    \small
    \begin{tabular}{l|lll|lll}
\toprule
 & \multicolumn{3}{l}{\textbf{AudioSet-50}} & \multicolumn{3}{l}{\textbf{ESC-50}} \\
 & \textit{AudioGen} & \textit{AudioLDM} &     \textit{CTAG} & \textit{AudioGen} & \textit{AudioLDM} &     \textit{CTAG} \\
\midrule
Complexity & 16.50 & 17.65 & 18.06 & 9.60 & 12.94 & 17.76 \\
      Flux & 0.08 & 0.11 & 0.18 & 0.06 & 0.09 & 0.15 \\
      HFC & 53.25 & 152.06 & 427.03 & 34.49 & 101.32 & 380.74 \\
      Rolloff & 2,487.71 & 1,628.55 & 7,031.67 & 2,254.98 & 1,647.51 & 6,996.19 \\
      Centroid & 1,629.95 & 1,096.16 & 4,139.99 & 1,512.55 & 1,108.42 & 4,227.08 \\
      Compression Ratio & 6.42 & 7.09 & 9.51 & 6.46 & 7.58 & 9.57 \\
\bottomrule
\end{tabular}
\caption{Comparison of spectral descriptors---complexity, flux, HFC, rolloff, centroid---and audio compression ratio, across ESC-50 and AudioSet-50. Results are grouped by the evaluation of three methods: \textit{AudioGen}, \textit{AudioLDM}, and our method, \textit{CTAG}.}
    \label{tab:results_q}
\end{table*}

We also report compression ratio, under variable bit rate (VBR) MP3 compression (quality = 4). Interestingly, \textit{CTAG} achieves a higher average compression ratio. VBR generally works by applying lower ratios to more perceptually complex input. Whether related to high-frequency content or other factors, this suggests \textit{CTAG} sounds contain more perceptual redundancy or are perceptually ``simpler''.

Note that none of these measures are validated as perceptual metrics of audio quality, and we do not intend to use them as such. Rather, they help us to quantify the qualitative differences we observe between \textit{CTAG}-synthesized sounds and other text-to-audio generation models' results.

\begin{table*}[!htb]
    \centering
    \small
    \begin{adjustbox}{width=\textwidth}
    \begin{tabular}{l|lllll|lllll}
\toprule
 & \multicolumn{5}{l}{\textbf{AudioSet-50}} & \multicolumn{5}{l}{\textbf{ESC-50}} \\
 & \textit{AudioGen} & \textit{AudioLDM} &     \textit{CTAG} & \textit{CTAG+T} & \textit{CTAG+C} & \textit{AudioGen} & \textit{AudioLDM} &   \textit{CTAG}    & \textit{CTAG+T} & \textit{CTAG+C}\\
\midrule
Acc (Top-1) &   51.6   &   17.4   &   26.2   &   25.2   &   23.6   &   54.0   &   23.0   &   16.4   &   11.4   &   13.8    \\
Acc (Top-5) &   77.4   &   44.2   &   45.2   &   52.2   &   51.6   &   71.8   &   49.4   &   30.4   &   26.4   &   31.0    \\
\bottomrule
\end{tabular}
\end{adjustbox}
\caption{Top-1 and Top-5 classification accuracies (\%) for pre-trained classifiers with AudioSet-50 and ESC-50. We evaluated both models on results collected using \textit{AudioGen}, \textit{AudioLDM}, and our method with just the class labels (\textit{CTAG}), a simple template (i.e. ``Sound of a ...'') for each sound (\textit{CTAG+T}) and finally using an LLM for prompt engineering (\textit{CTAG+C}).}
    \label{tab:results_c}
\end{table*}

\subsection{User Study}
We recruited 10 participants via Prolific at \$12/h for a total of \$53.33, resulting in a total of 600 observations per outcome variable (i.e. accuracy, confidence, and artistic interpretiveness). \Cref{tab:results_survey} contains the results, which show that our sounds were identified by listeners substantially more accurately than those from \textit{AudioLDM} (odds ratio = 2.72, 95\% CI [1.61, 4.58], $p<.0001$), and only slightly less than \textit{AudioGen} on average (odds ratio = 0.85, 95\% CI [0.51, 1.42], $p=1$). Interestingly, though the confidence ratings replicate the ordering of the accuracy results, respondents were significantly more confident rating \textit{AudioGen} sounds, and reported similar, lower confidence levels for both \textit{CTAG} and \textit{AudioLDM}. This underscores the abstractness of \textit{CTAG}'s sounds; despite being identified more correctly, they still create uncertainty.

\begin{table}[!htb]
    \centering
    \small
\begin{tabular}{l|lll}
\toprule
 & \textit{AudioGen} & \textit{AudioLDM} & \textit{CTAG} \\
\midrule
Accuracy                &    59.5 &   34.0 &   56.0 \\
Confidence              &    3.48 &   2.95 &   2.99 \\
Artistic Interpretation &    2.32 &   2.90 &   3.54 \\
\bottomrule
\end{tabular}
\caption{User study results for sounds from \textit{AudioGen}, \textit{AudioLDM}, and our method, \textit{CTAG}. We report accuracy percentage and confidence (1–5) on label identification, and average rating of the artistic interpretiveness (1–5) of the sound. Overall, \textit{CTAG} retains competitive identifiability while being perceived as more artistic.}
\label{tab:results_survey}
\end{table}

Results also show \textit{CTAG} sounds were perceived to be significantly more artistically interpretive than both \textit{AudioGen} (contrast = 1.22, 95\% CI [0.93, 1.51], $t(579) = 10.20$, $p<.0001$) and \textit{AudioLDM} (contrast = 0.65, 95\% CI [0.36, 0.93], $t(579) = 5.39$, $p<.0001$).

These findings highlight our approach's benefits in capturing artistic interpretation compared to both the existing approaches. All $p$-values are Bonferroni-adjusted. Full results for post-hoc contrasts are available in the Appendix.

\subsection{Additional Analyses}
In \Cref{sec:supplementary_analyses} we provide additional analyses relating to generation time, CLAP scores, prompting strategies for the baseline models, user study results, and a visualization of the parameter space of \textit{CTAG}-generated sounds.

\section{Limitations}
Our method requires iterating for each prompt from random initialization, but techniques like semantic caching to initialize to similar prompts' parameters, predictive methods for prompt-to-parameter derivation, and a user interface extension for tweaking parameters are all potential extensions to make our method more useful in real-world settings. We also focus on brief, non-mixture sounds as these are what the synthesizer is suited to modeling. Future work could explore strategies to extend the duration and complexity of sounds that can be synthesized this way.

\section{Conclusion}
In this work, we proposed a method for text-to-audio generation that offers a fresh perspective on neural audio synthesis by using a virtual modular synthesizer. This approach emphasizes the meaningful abstraction of auditory phenomena, contrary to prevalent methods that prioritize acoustic realism. Our results position this approach as a distinctive tool in the field of audio synthesis, capable of both expanding the toolkit of novices and experts, and stimulating new directions in audio generation research.

\section*{Acknowledgements}
The authors acknowledge the MIT SuperCloud and Lincoln Laboratory Supercomputing Center for providing resources that have contributed to the research results reported within this paper. Manuel was supported by a US-Spain Fulbright grant. We extend our heartfelt thanks to all participants in the user study. We also thank our meta-reviewer and reviewers, as well as Yingtao Tian, Ashvala Vinay, and Hugo Flores García for supportive comments.

\section*{Impact Statement}
This work introduces a novel method for generating abstract and creative sounds from text prompts, intending to expand the creative possibilities for text-to-audio generation. We foresee several potentially positive societal impacts: (1) democratizing access to creative sound design tools, (2) stimulating new directions in audio machine learning research, (3) personalization and customization, (4) lowered likelihood of re-generating training data, and (5) lowering the computation barrier. 

We do not foresee direct negative societal consequences from this contribution. However, as with any generative technology, there exists potential for misuse which should be monitored. For example, synthesized sounds are not always identifiable, and should not be used in high-stakes circumstances where identification is essential. Additionally, synthesizers can simplify complex real-world phenomena; we recognize sounds can convey a rich variety of information beyond this.

\bibliography{main}
\bibliographystyle{icml2024}

\newpage
\appendix
\onecolumn

\section{Supplementary Analyses}
\label{sec:supplementary_analyses}

\subsection{Generation Time}

\begin{table}[!htb]
\centering
\begin{tabular}{lrrr}
\toprule
\textbf{Iter/Popsize} & \textbf{25} & \textbf{50} & \textbf{100} \\
\midrule
50 & 5.49 $\pm$ 0.154 & 9.62 $\pm$ 0.452 & 18.43 $\pm$ 0.752 \\
100 & 10.01 $\pm$ 0.194 & 18.05 $\pm$ 0.605 & 33.40 $\pm$ 0.331 \\
300 & 27.61 $\pm$ 0.703 & 49.94 $\pm$ 0.424 & 97.23 $\pm$ 0.469 \\
\bottomrule
\end{tabular}
\caption{Time (in seconds) for different population sizes (columns) and iteration counts (rows).}
\label{tab:times}
\end{table}

In \Cref{tab:times} we illustrate the optimization times, in seconds, for different numbers of iterations (rows) and optimizer population sizes (columns) below, on a modest GPU, i.e. single V100. Note that the necessary number of iterations varies for different prompts, from 50 to 300+ to get optimal results.

\subsection{CLAP Scores}
\begin{table}[!htb]
\centering
\begin{tabular}{lrr}
\toprule
\textbf{Model} & \textbf{AudioSet-50} & \textbf{ESC-50} \\
\midrule
\textit{AudioGen} & 0.249 $\pm$ 0.160 & 0.277 $\pm$ 0.180 \\
\textit{AudioLDM} & 0.166 $\pm$ 0.128 & 0.173 $\pm$ 0.142 \\
\textit{CTAG} & \textbf{0.573} $\pm$ 0.126 & \textbf{0.585} $\pm$ 0.130 \\
Real & -- & 0.416 $\pm$ 0.139 \\
\bottomrule
\end{tabular}
\caption{Comparison of CLAP scores between \textit{CTAG} and other generative models on AudioSet-50 and ESC-50 datasets}
\label{tab:clap_scores}
\end{table}

\Cref{tab:clap_scores} shows the CLAP~\cite{wu2023large} evaluations for each model with AudioSet-50 and ESC-50 prompts, as well as for the actual ESC-50 dataset of real sounds. CLAP is the objective that we optimize in our synthesis-by-optimization approach, and these results show how \textit{CTAG} trivially achieves a higher score compared to all other models and even the real data. This highlights the ability of our optimization strategy to effectively maximize the CLAP score, and also the importance of finding alternative and distinct evaluation metrics as we showed in \Cref{sec:eval_metrics}.

\subsection{Prompting Strategies for All Tested Models}
\begin{table}[!htb]
\centering
\begin{tabular}{lllrrr}
\toprule
\textbf{Dataset} & \textbf{Metric} & \textbf{Model} & \textbf{Sounds} & \textbf{Template} & \textbf{Caption} \\
\midrule
\multirow{6}{*}{AudioSet-50} & \multirow{3}{*}{Top-1} & \textit{AudioGen} & 51.6 & \textbf{57.0} & 48.8 \\
 &  & \textit{AudioLDM} & 17.4 & \textbf{21.0} & 16.6 \\
 &  & \textit{CTAG} & \textbf{26.2} & 25.2 & 23.6 \\
\cline{2-6}
 & \multirow{3}{*}{Top-5} & \textit{AudioGen} & 77.4 & \textbf{84.8} & 80.8 \\
 &  & \textit{AudioLDM} & 44.2 & \textbf{49.8} & 48.0 \\
 &  & \textit{CTAG} & 45.2 & \textbf{52.2} & 51.6 \\
\midrule
\multirow{6}{*}{ESC-50} & \multirow{3}{*}{Top-1} & \textit{AudioGen} & 54.0 & \textbf{69.0} & 62.0 \\
 &  & \textit{AudioLDM} & 23.0 & 20.2 & \textbf{29.4} \\
 &  & \textit{CTAG} & \textbf{16.4} & 11.4 & 13.8 \\
\cline{2-6}
 & \multirow{3}{*}{Top-5} & \textit{AudioGen} & 71.8 & \textbf{85.2} & 81.8 \\
 &  & \textit{AudioLDM} & 49.4 & 47.0 & \textbf{58.4} \\
 &  & \textit{CTAG} & 30.2 & 26.4 & \textbf{31.0} \\
\bottomrule
\end{tabular}
\caption{Performance comparison, with different prompting strategies, of models on AudioSet-50 and ESC-50 datasets}
\label{tab:prompting_strategies_all_models}
\end{table}

For completeness, \Cref{tab:prompting_strategies_all_models} provides all the results for all different models with templates and captions as we showed for \textit{CTAG} in \Cref{sec:classification_results}. The performance of \textit{AudioGen} shows a notable boost when using the +T (Template) strategy. However, the impact of these strategies on the other models and datasets is less consistent, with some cases showing modest improvements and others exhibiting a decrease in performance (e.g., \textit{AudioLDM} ESC-50 +T, \textit{AudioLDM} AudioSet-50 +C). Given the variability in results, it is difficult to make a definitive statement about the effectiveness of these strategies across all baselines. While they may prove beneficial in certain scenarios, their impact appears to be context-dependent.

\subsection{User Study Statistical Models}
We report post-hoc contrasts for the user study results in \Cref{tab:accuracy,tab:confidence,tab:interpretation}.

\begin{table}[!htb]
\centering
\centering
\begin{tabular}[t]{lrrrrrl}
\toprule
contrast & odds.ratio & SE & asymp.LCL & asymp.UCL & z.ratio & p.value\\
\midrule
\textit{AudioLDM} / \textit{AudioGen} & 0.31 & 0.07 & 0.19 & 0.53 & -5.28 & $<$1e-04\\
\textit{CTAG} / \textit{AudioGen} & 0.85 & 0.18 & 0.51 & 1.42 & -0.75 & 1\\
\textit{CTAG} / \textit{AudioLDM} & 2.72 & 0.59 & 1.61 & 4.58 & 4.59 & $<$1e-04\\
\bottomrule
\end{tabular}
\caption{Post-hoc contrasts from a mixed-effects logistic regression for accuracy.}
\label{tab:accuracy}
\end{table}

\begin{table}[!htb]
\centering
\centering
\begin{tabular}[t]{lrrrrrrl}
\toprule
contrast & estimate & SE & df & lower.CL & upper.CL & t.ratio & p.value\\
\midrule
\textit{AudioLDM} - \textit{AudioGen} & -0.53 & 0.12 & 579 & -0.82 & -0.24 & -4.34 & $<$1e-04\\
\textit{CTAG} - \textit{AudioGen} & -0.48 & 0.12 & 579 & -0.78 & -0.19 & -3.97 & 0.00024\\
\textit{CTAG} - \textit{AudioLDM} & 0.04 & 0.12 & 579 & -0.25 & 0.34 & 0.37 & 1\\
\bottomrule
\end{tabular}
\caption{Post-hoc contrasts from a mixed-effects linear regression for confidence ratings.}
\label{tab:confidence}
\end{table}

\begin{table}[!htb]
\centering
\centering
\begin{tabular}[t]{lrrrrrrl}
\toprule
contrast & estimate & SE & df & lower.CL & upper.CL & t.ratio & p.value\\
\midrule
\textit{AudioLDM} - \textit{AudioGen} & 0.57 & 0.12 & 579 & 0.29 & 0.86 & 4.81 & $<$1e-04\\
\textit{CTAG} - \textit{AudioGen} & 1.22 & 0.12 & 579 & 0.93 & 1.51 & 10.20 & $<$1e-04\\
\textit{CTAG} - \textit{AudioLDM} & 0.65 & 0.12 & 579 & 0.36 & 0.93 & 5.39 & $<$1e-04\\
\bottomrule
\end{tabular}
\caption{Post-hoc contrasts from a mixed-effects linear regression for artistic interpretativeness.}
\label{tab:interpretation}
\end{table}

\subsection{User Study Per-Prompt Accuracy}
\begin{figure}[!htb]
    \centering
    \includegraphics[width=0.68\textwidth]{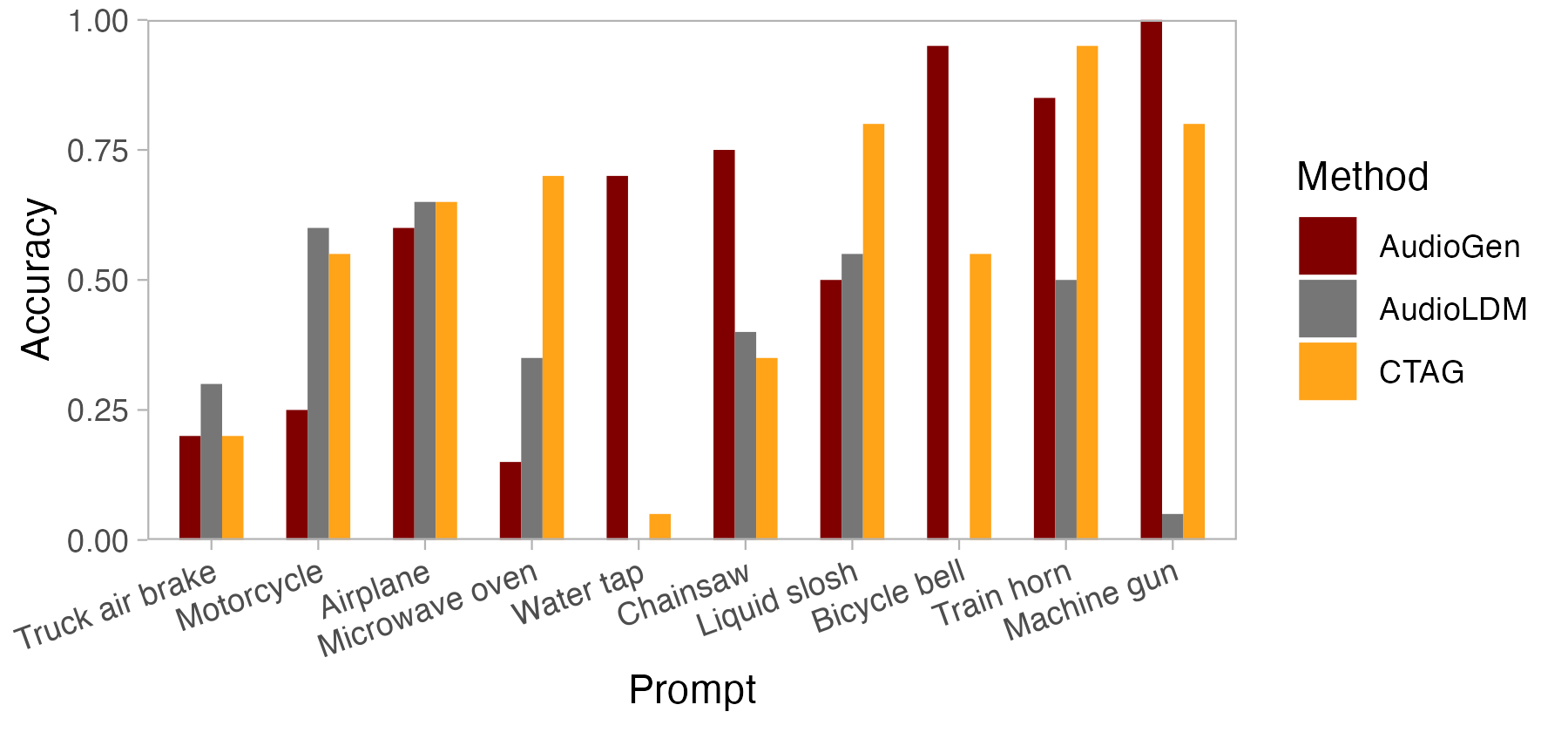}
    \caption{User study classification accuracy per prompt, for \textit{CTAG}, \textit{AudioGen}, and \textit{AudioLDM}.}
    \label{fig:accuracy_prompt}
\end{figure}

\Cref{fig:accuracy_prompt} shows the accuracy of our user study participants at classifying sounds generated with \textit{CTAG}, \textit{AudioGen}, and \textit{AudioLDM}. Reviewing these differences shows that some sounds are overall more difficult to identify, for instance; ``Truck air brake''. This may be due to the ambiguity in what this can sound like, as it is not as common a sound as ``Bicycle bell''.

\subsection{Dimensionality Reduction}
\begin{figure}[t]
\centering
\includegraphics[width=0.8\textwidth]{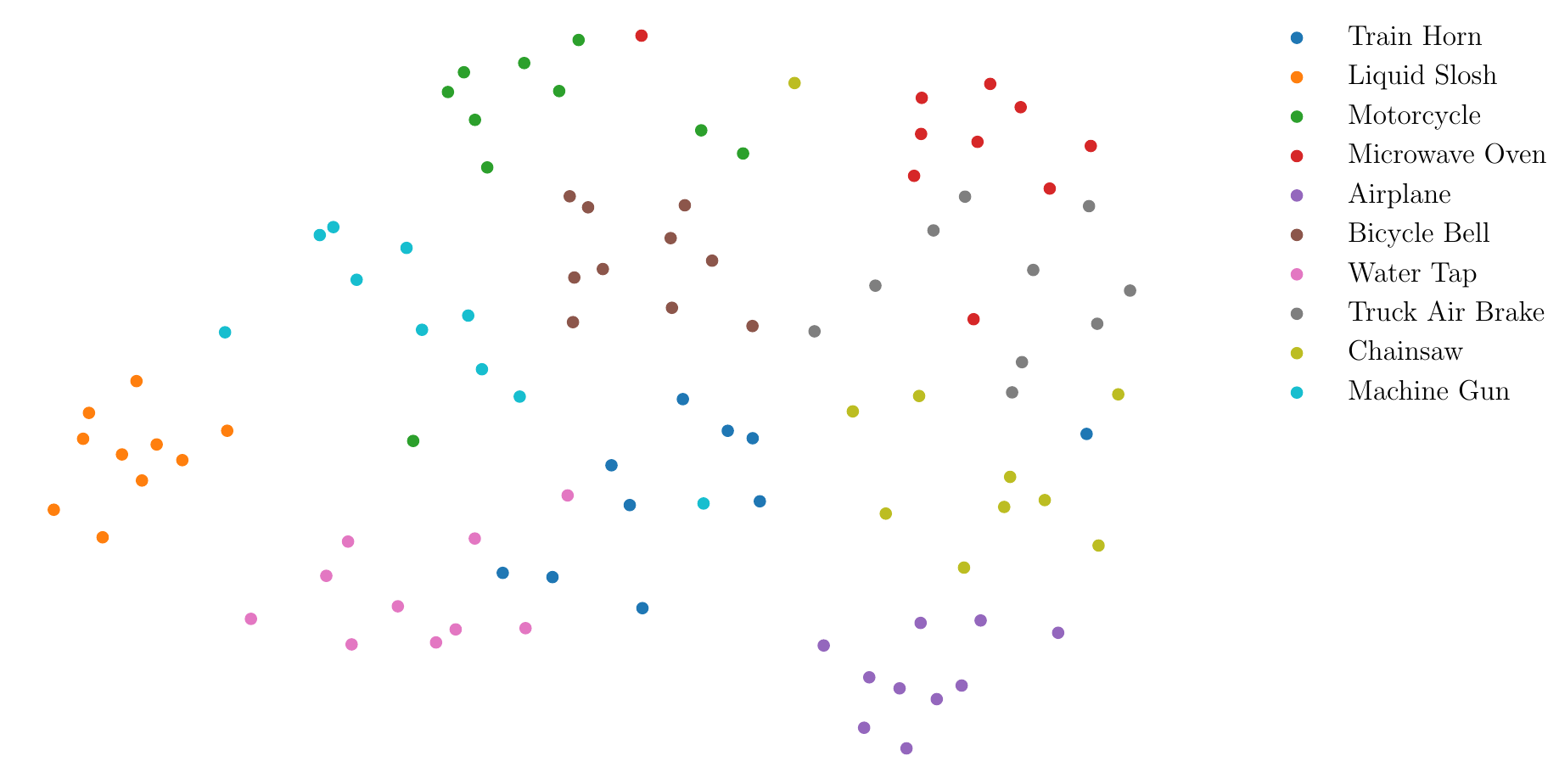}
\caption{Dimensionality reduction of the \textit{Voice} synthesizer parameters using UMAP applied to 10 sounds from each of the 10 classes from the user study. I   t distinctly reveals clusters corresponding to individual sounds, and it shows how conceptually similar sounds such as ``water tap'' and ``liquid slosh'' are closer in space.}
\label{fig:umap}
\end{figure}

Having access to the parameters of the synthesizer also allows us to project them into a two-dimensional space to explore the relationship between sounds. Leveraging the Uniform Manifold Approximation and Projection (UMAP) \cite{mcinnes2018umap} algorithm for dimensionality reduction of the synthesizer parameters, \Cref{fig:umap} shows how the representation delineates clusters for each distinct sound class while retaining semantic meaning---sounds with similar acoustic properties cluster together.

\section{Caption Prompt}
We used the following instructions to generate caption-like prompts from class labels:
\begin{quote}
\textit{``Write a simple one-sentence audio caption that describes objectively each sound itself in a real scenario without making up any extra details about other possible sounds or places. You should define the most common action for such an entity when multiple options are available. Avoid using templates such as `A sound of' or `The sound of'. Sounds: [List]''}
\end{quote}

\section{Listener Survey}
In this section, we provide information about the survey design we used to collect human ratings.

\subsection{Survey Flow}
\begin{itemize}
    \item Standard: Introduction (3 Questions)
    \item Block: Audio (4 Questions)
    \item Standard: Additional (2 Questions)
\end{itemize}

\subsection{Start of Block: Introduction}
\textbf{Q1}: We are conducting a survey to assess the quality of a novel method for text-to-audio generation. You will be presented with a series of short sounds, and asked to select the closest category from a given list, the confidence in your prediction, and how artistically designed the sound is compared to a more realistic interpretation.

\noindent\textbf{Q2}: I consent to participate. I understand that my participation is voluntary and I may withdraw my consent at any time.
\begin{itemize}
    \item Yes  (1) 
    \item No (2)
\end{itemize}

\noindent\textbf{Q3}: I am at least 18 years old.
\begin{itemize}
    \item Yes  (1) 
    \item No (2) 
\end{itemize}

\noindent\textbf{Q4}: Do you have any hearing loss or hearing difficulties?
\begin{itemize}
    \item Yes  (1) 
    \item No (2) 
\end{itemize}

\noindent\textbf{Q5}: Are you fluent in English?
\begin{itemize}
    \item Yes  (1) 
    \item No (2) 
\end{itemize}

\noindent\textbf{Q5}: What is your Prolific ID?
Please note that this response should auto-fill with the correct ID

\subsection{Start of Block: Audio}

We use Qualtrics' Loop \& Merge functionality to loop through the sounds.

\noindent\textbf{A}: Select the closest category for the following sound:
\textbf{[Audio Clip]}
\begin{itemize}
    \item Truck air brake (1)
    \item Water tap (2)
    \item Train horn (3)
    \item Motorcycle (4)
    \item Microwave oven (5)
    \item Liquid slosh (6)
    \item Chainsaw (7)
    \item Airplane (8)
    \item Bicycle bell (9)
    \item Machine gun (10)
\end{itemize}

\noindent\textbf{B}: How confident are you in your selected answer?
\begin{itemize}
    \item Completely confident  (1)
    \item Fairly confident  (2)
    \item Somewhat confident  (3)
    \item Slightly confident  (4)
    \item Not confident at all  (5)
\end{itemize}

\noindent\textbf{C}: Would you associate this sound more with a realistic portrayal or an artistic interpretation of the category that you selected?
\begin{itemize}
    \item 1 (1) Realistic Portrayal
    \item 2 (2) •
    \item 3 (3) •
    \item 4 (4) •
    \item 5 (5) Artistic Interpretation
\end{itemize}

\subsection{Start of Block: Additional}
We have two questions to check that participants were paying attention.

\noindent\textbf{A1} Please select "Chainsaw" from the options below:
\begin{itemize}
    \item Truck air brake (1)
    \item Water tap (2)
    \item Train horn (3)
    \item Motorcycle (4)
    \item Microwave oven (5)
    \item Liquid slosh (6)
    \item Chainsaw (7)
    \item Airplane (8)
    \item Bicycle bell (9)
    \item Machine gun (10)
\end{itemize}

\noindent\textbf{A2}: All of the sounds you heard during this survey were the same.
\begin{itemize}
    \item Yes  (1)
    \item No  (2)
\end{itemize}

\noindent\textbf{Completion Message}: Thank you for taking part in this study. Please click the button below to be redirected back to Prolific and register your submission.


\end{document}